\input harvmac
\def\figflag{I}
\noblackbox

\font\cmss=cmss10 \font\cmsss=cmss10 at 7pt
 \def\inbar{\,\vrule height1.5ex width.4pt depth0pt}
\def\IZ{\relax\ifmmode\mathchoice
{\hbox{\cmss Z\kern-.4em Z}}{\hbox{\cmss Z\kern-.4em Z}}
{\lower.9pt\hbox{\cmsss Z\kern-.4em Z}}
{\lower1.2pt\hbox{\cmsss Z\kern-.4em Z}}\else{\cmss Z\kern-.4em
Z}\fi}
\def\IB{\relax{\rm I\kern-.18em B}}
\def\IC{{\relax\hbox{$\inbar\kern-.3em{\rm C}$}}}
\def\ID{\relax{\rm I\kern-.18em D}}
\def\IE{\relax{\rm I\kern-.18em E}}
\def\IF{\relax{\rm I\kern-.18em F}}
\def\IG{\relax\hbox{$\inbar\kern-.3em{\rm G}$}}
\def\IGa{\relax\hbox{${\rm I}\kern-.18em\Gamma$}}
\def\IH{\relax{\rm I\kern-.18em H}}
\def\II{\relax{\rm I\kern-.18em I}}
\def\IK{\relax{\rm I\kern-.18em K}}
\def\IR{\relax{\rm I\kern-.18em R}}

\def\tfig#1{Figure~\the\figno\xdef#1{Figure~\the\figno}\global\advance\figno by1}
\def\figI{I}
%
\newdimen\tempszb \newdimen\tempszc \newdimen\tempszd \newdimen\tempsze
\ifx\figflag\figI
\input epsf
%
%
%
%
%
\def\ifigure#1#2#3#4{
\midinsert
\vbox to #4truein{\ifx\figflag\figI
\vfil\centerline{\epsfysize=#4truein\epsfbox{#3}}\fi}
\baselineskip=12pt
\narrower\narrower\noindent{\bf #1:} #2
\endinsert
}

\lref\FloreaSI{
  B.~Florea, S.~Kachru, J.~McGreevy and N.~Saulina,
    ``Stringy instantons and quiver gauge theories,''
      JHEP {\bf 0705}, 024 (2007)
        [arXiv:hep-th/0610003].
      }
\lref\BlumenhagenXT{
  R.~Blumenhagen, M.~Cvetic and T.~Weigand,
    ``Spacetime instanton corrections in 4D string vacua - the seesaw mechanism
      for D-brane models,''
        Nucl.\ Phys.\  B {\bf 771}, 113 (2007)
      [arXiv:hep-th/0609191].
        }
\lref\IntriligatorDD{
  K.~Intriligator, N.~Seiberg and D.~Shih,
    ``Dynamical SUSY breaking in meta-stable vacua,''
      JHEP {\bf 0604}, 021 (2006)
        [arXiv:hep-th/0602239].
      }
\lref\CveticKU{
  M.~Cvetic, R.~Richter and T.~Weigand,
    ``Computation of D-brane instanton induced superpotential couplings -
      Majorana masses from string theory,''
        arXiv:hep-th/0703028.
      }
\lref\AbelYK{
  S.~A.~Abel and M.~D.~Goodsell,
    ``Realistic Yukawa couplings through instantons in intersecting brane
      worlds,''
        arXiv:hep-th/0612110.
      }
\lref\KitanoXG{
  R.~Kitano, H.~Ooguri and Y.~Ookouchi,
    ``Direct mediation of meta-stable supersymmetry breaking,''
      Phys.\ Rev.\  D {\bf 75}, 045022 (2007)
        [arXiv:hep-ph/0612139].
      }
\lref\BlumenhagenZK{
  R.~Blumenhagen, M.~Cvetic, D.~Lust, R.~Richter and T.~Weigand,
    ``Non-perturbative Yukawa Couplings from String Instantons,''
      arXiv:0707.1871 [hep-th].
    }
\lref\AkerblomUC{
  N.~Akerblom, R.~Blumenhagen, D.~Lust and M.~Schmidt-Sommerfeld,
    ``Instantons and Holomorphic Couplings in Intersecting D-brane Models,''
      arXiv:0705.2366 [hep-th].
    }
\lref\AkerblomHX{
  N.~Akerblom, R.~Blumenhagen, D.~Lust, E.~Plauschinn and M.~Schmidt-Sommerfeld,
    ``Non-perturbative SQCD Superpotentials from String Instantons,''
      JHEP {\bf 0704}, 076 (2007)
        [arXiv:hep-th/0612132].
      }
\lref\AntuschJD{
  S.~Antusch, L.~E.~Ibanez and T.~Macri,
    ``Neutrino Masses and Mixings from String Theory Instantons,''
      arXiv:0706.2132 [hep-ph].
    }
\lref\IbanezRS{
  L.~E.~Ibanez, A.~N.~Schellekens and A.~M.~Uranga,
    ``Instanton Induced Neutrino Majorana Masses in CFT Orientifolds with
      MSSM-like spectra,''
        arXiv:0704.1079 [hep-th].
      }
\lref\DiaconescuAH{
  D.~E.~Diaconescu, R.~Donagi and B.~Florea,
      ``Metastable quivers in string compactifications,''
      Nucl.\ Phys.\  B {\bf 774}, 102 (2007)
        [arXiv:hep-th/0701104].
      }
\lref\IbanezDA{
  L.~E.~Ibanez and A.~M.~Uranga,
    ``Neutrino Majorana masses from string theory instanton effects,''
      JHEP {\bf 0703}, 052 (2007)
        [arXiv:hep-th/0609213].}
\lref\HaackCY{
  M.~Haack, D.~Krefl, D.~Lust, A.~Van Proeyen and M.~Zagermann,
    ``Gaugino condensates and D-terms from D7-branes,''
      JHEP {\bf 0701}, 078 (2007)
        [arXiv:hep-th/0609211].
      }
\lref\FrancoII{
  S.~Franco, A.~Hanany, D.~Krefl, J.~Park, A.~M.~Uranga and D.~Vegh,
    ``Dimers and Orientifolds,''
      arXiv:0707.0298 [hep-th].
    }
\lref\BianchiWY{
  M.~Bianchi, F.~Fucito and J.~F.~Morales,
    ``D-brane Instantons on the $T^6/Z_3$ orientifold,''
      arXiv:0704.0784 [hep-th].
    }
\lref\BilloKQ{
  M.~Billo, M.~Frau and A.~Lerda,
  ``N=2 Instanton Calculus In Closed String Background,''
  arXiv:0707.2298 [hep-th].
}

    \lref\BianchiFX{
      M.~Bianchi and E.~Kiritsis.
      ``Non-perturbative and Flux superpotentials for Type I strings on the $Z_3$ orbifold,''
            arXiv:hep-th/0702015.
          }
\lref\ArgurioVQ{
  R.~Argurio, M.~Bertolini, G.~Ferretti, A.~Lerda and C.~Petersson,
    ``Stringy Instantons at Orbifold Singularities,''
      arXiv:0704.0262 [hep-th].
    }

\lref\AspinwallBS{
  P.~S.~Aspinwall and S.~H.~Katz,
  ``Computation of superpotentials for D-Branes,''
  Commun.\ Math.\ Phys.\  {\bf 264}, 227 (2006)
  [arXiv:hep-th/0412209].
}

\lref\ArgurioQK{
  R.~Argurio, M.~Bertolini, S.~Franco and S.~Kachru,
    ``Metastable vacua and D-branes at the conifold,''
      JHEP {\bf 0706}, 017 (2007)
        [arXiv:hep-th/0703236].
      }
\lref\ArgurioNY{
  R.~Argurio, M.~Bertolini, S.~Franco and S.~Kachru,
    ``Gauge/gravity duality and meta-stable dynamical supersymmetry breaking,''
      JHEP {\bf 0701}, 083 (2007)
        [arXiv:hep-th/0610212].
      }
            \lref\KlebanovHB{
                I.~R.~Klebanov and M.~J.~Strassler,
``Supergravity and a confining gauge theory: Duality cascades and
$\chi$SB-resolution of naked singularities,'' JHEP {\bf 0008}, 052
(2000) [arXiv:hep-th/0007191].
}
\lref\GanorPE{
  O.~J.~Ganor,
    ``A note on zeroes of superpotentials in F-theory,''
      Nucl.\ Phys.\  B {\bf 499}, 55 (1997)
        [arXiv:hep-th/9612077].
      }
\lref\DouglasSW{
  M.~R.~Douglas and G.~W.~Moore,
    ``D-branes, Quivers, and ALE Instantons,''
      arXiv:hep-th/9603167.
    }
\lref\AharonyZR{
  O.~Aharony, A.~Buchel and A.~Yarom,
  ``Holographic renormalization of cascading gauge theories,''
  Phys.\ Rev.\  D {\bf 72}, 066003 (2005)
  [arXiv:hep-th/0506002].
}

\lref\UrangaVF{
  A.~M.~Uranga,
    ``Brane configurations for branes at conifolds,''
      JHEP {\bf 9901}, 022 (1999)
        [arXiv:hep-th/9811004].
      }
\lref\WittenBN{
  E.~Witten,
    ``Non-Perturbative Superpotentials In String Theory,''
      Nucl.\ Phys.\  B {\bf 474}, 343 (1996)
        [arXiv:hep-th/9604030].
      }
\lref\KachruYS{
  S.~Kachru and E.~Silverstein,
    ``4d conformal theories and strings on orbifolds,''
      Phys.\ Rev.\ Lett.\  {\bf 80}, 4855 (1998)
        [arXiv:hep-th/9802183].
      }
\lref\KlebanovHH{
  I.~R.~Klebanov and E.~Witten,
    ``Superconformal field theory on threebranes at a Calabi-Yau  singularity,''
      Nucl.\ Phys.\  B {\bf 536}, 199 (1998)
        [arXiv:hep-th/9807080].
      }
\lref\GorlichQM{
  L.~Gorlich, S.~Kachru, P.~K.~Tripathy and S.~P.~Trivedi,
    ``Gaugino condensation and nonperturbative superpotentials in flux
      compactifications,''
        JHEP {\bf 0412}, 074 (2004)
      [arXiv:hep-th/0407130].
        }
\lref\KalloshGS{
  R.~Kallosh, A.~K.~Kashani-Poor and A.~Tomasiello,
    ``Counting fermionic zero modes on M5 with fluxes,''
      JHEP {\bf 0506}, 069 (2005)
        [arXiv:hep-th/0503138].
      }
\lref\BergshoeffYP{
  E.~Bergshoeff, R.~Kallosh, A.~K.~Kashani-Poor, D.~Sorokin and A.~Tomasiello,
    ``An index for the Dirac operator on D3 branes with background fluxes,''
      JHEP {\bf 0510}, 102 (2005)
        [arXiv:hep-th/0507069].
      }
\lref\LustCU{
  D.~Lust, S.~Reffert, W.~Schulgin and P.~K.~Tripathy,
    ``Fermion zero modes in the presence of fluxes and a non-perturbative
      superpotential,''
        JHEP {\bf 0608}, 071 (2006)
      [arXiv:hep-th/0509082].
        }
\lref\SaulinaVE{
  N.~Saulina,
    ``Topological constraints on stabilized flux vacua,''
      Nucl.\ Phys.\  B {\bf 720}, 203 (2005)
        [arXiv:hep-th/0503125].
      }
\lref\LawrenceJA{
  A.~E.~Lawrence, N.~Nekrasov and C.~Vafa,
    ``On conformal field theories in four dimensions,''
      Nucl.\ Phys.\  B {\bf 533}, 199 (1998)
        [arXiv:hep-th/9803015].
      }
\lref\TripathyHV{
  P.~K.~Tripathy and S.~P.~Trivedi,
    ``D3 brane action and fermion zero modes in presence of background flux,''
      JHEP {\bf 0506}, 066 (2005)
        [arXiv:hep-th/0503072].
      }


\Title {\vbox{\baselineskip12pt\hbox{SLAC-PUB-12680}
\hbox{SU-ITP-07/09} \hbox{WIS/10/07-JUL-DPP}}} {\vbox{\centerline{
Stringy Instantons and Cascading Quivers} } } \centerline{Ofer
Aharony$^{1,2}$ and Shamit Kachru$^{2}$ }

\bigskip
\centerline{$^{1}$Department of Particle Physics}
\centerline{Weizmann Institute of Science}
\centerline{Rehovot 76100, Israel}
\smallskip
\medskip
\centerline{$^{2}$ Department of Physics and SLAC}
\centerline{Stanford University}
\centerline{Stanford, CA 94305, USA}
\medskip
\bigskip
\noindent D-brane instantons can perturb the quantum field theories
on space-time filling D-branes by interesting operators. In some
cases, these D-brane instantons are novel ``stringy'' effects (not
interpretable directly as instanton effects in the low-energy
quantum field theory), while in others the D-brane instantons can be
directly interpreted as field theory effects. In this note, we
describe a situation where both perspectives are available, by
studying stringy instantons in quivers which arise at simple
Calabi-Yau singularities. We show that a stringy instanton which
wraps an unoccupied node of the quiver, and gives rise to a
non-perturbative mass in the space-time field theory, can be
reinterpreted as a conventional gauge theory effect by going up in
an appropriate renormalization group cascade. Interestingly, in the
cascade, the contribution of the stringy instanton does not come
from gauge theory instantons but from strong coupling dynamics.

\Date{July 2007}


\newsec{Introduction}

Quantum effects which are non-perturbatively small in the coupling
constant $g$ may play an important role in many physical
phenomena. For instance, they may be relevant to dynamical
(super)symmetry breaking, or provide a natural mechanism to
generate small Yukawa couplings or masses in the Lagrangian.  In
string theory, instantons which generate such effects can often be
geometrized as D-branes.  Investigation of both novel stringy
effects (not obviously interpretable as field theory instanton
effects) \refs{\BlumenhagenXT,\IbanezDA,\FloreaSI} and more
conventional field theory instanton effects
\refs{\HaackCY,\FloreaSI,\AkerblomHX} involving Euclidean D-branes
has recently been initiated by several groups (building in part on
important earlier work of Ganor \GanorPE\ and Witten \WittenBN).
Further explorations and applications of these instantons appear
in \refs{\AbelYK\DiaconescuAH
\BianchiFX\CveticKU\ArgurioQK\ArgurioVQ\BianchiWY\IbanezRS
\AkerblomUC\AntuschJD\FrancoII\BlumenhagenZK-\BilloKQ}.

A rich set of theories where such effects may be computable and
important is provided by D-branes at Calabi-Yau singularities. Our
goal in this paper will be to use such a system to demonstrate a
D-brane instanton effect in two different and complementary ways.
Branes at singularities give rise rather generally to quiver gauge
theories \DouglasSW. If some nodes of the quiver are unoccupied by
space-filling branes, one may still construct interesting
instantons by wrapping Euclidean branes over the corresponding
cycles in the geometry \FloreaSI.  These can give rise to stringy
perturbations of the low-energy effective theory, which are not
directly interpreted as instanton effects in the quiver gauge
theory.

On the other hand, many quivers corresponding to four dimensional
${\cal N}=1$ supersymmetric gauge theories exhibit renormalization
group (RG) cascades \KlebanovHB. These include D-branes at
conifolds and their generalizations (but not orbifold
singularities, which have free worldsheet descriptions). The
cascades describe brane systems where all nodes of the quiver are
occupied at high energies in the field theory, while the low
energy physics may be described by a brane configuration with some
unoccupied nodes.

In orbifolds, one can directly use worldsheet techniques to
systematically derive the string instanton contributions. While
this is not possible for conifolds and their generalizations, in
such systems, one instead has the intriguing possibility of
computing the D-brane instanton effect generated by a brane
wrapping an unoccupied quiver node, in two different ways:
\medskip
\noindent 1) One can do the path integral over D-instanton
collective coordinates in the quiver with occupation numbers
describing the end of the cascade.  This requires use of
sophisticated mathematical technology to infer the action on the
D-instanton \AspinwallBS.
\medskip
\noindent 2) One can try to derive the ${\it same}$ effect by
analyzing the gauge theory at higher steps in the cascade, where
the relevant node is occupied by space-filling branes.  In this
case
one should be able to reproduce the desired
effect by using standard techniques in ${\cal N}=1$ supersymmetric
gauge theory.

\noindent The agreement between methods 1) and 2) that we find (in a
particular simple system that we can analyze in detail), can be
viewed as a consistency check on the presence and form of the novel
stringy effects. Somewhat surprisingly, in the second method we find
that the effect does not come from field theory instantons but from
strong coupling dynamics.

The agreement between the two methods is intuitively expected for
the following reason.  The gauge theory at higher steps in the
cascade in method 2) can be UV completed by embedding an
appropriate brane configuration in a non-compact, singular
Calabi-Yau manifold\foot{The cascade does not necessarily need to
be completed in this manner. The cascade with an infinite number
of steps can be defined by the holographic renormalization of its
gravity dual \AharonyZR.}. We can also do this directly with the
partially occupied quiver of method 1). The superpotential of the
low-energy field theory is not expected to depend on the number of
cascade steps $K$. As we lower $K$, eventually the gauge theory
effect that we compute by method 2), becomes a stringy effect that
we compute by method 1).

The organization of this paper is as follows.  In \S2, we introduce
the singularity and the IR brane configuration that we will study.
This configuration was previously explored in
\refs{\ArgurioNY,\ArgurioQK}. In \S3, we review the expected stringy
instanton effect. In \S4, we describe an RG cascade which ends with
this brane configuration. In \S5, we show that careful analysis of
the field theory dynamics along the cascade reproduces the expected
result of \S3.

Although our analysis is limited to one illustrative case, we
expect that similar results could be obtained for instanton
computations in quivers characterizing more general Calabi-Yau
singularities.

\newsec{The quiver of branes at orbifolds of the conifold}

Branes at singularities provide interesting gauge theories which
exhibit reduced supersymmetry and intricate (non-conformal) IR
dynamics, and which in many cases admit dual gravity descriptions.
The simplest cases involve orbifold
\refs{\DouglasSW,\KachruYS,\LawrenceJA} and conifold \KlebanovHH\
singularities.

The conifold can be described by the equation \eqn\conis{ xy = zw
} in $\IC^4$.  The quiver which captures the field theory on
D3-branes and fractional D5-branes at this singularity, in type
IIB string theory, is shown in \tfig\figone.

\ifigure\figone{Quiver diagram for the conifold singularity. Each
node is an $SU(r_i)$ gauge group, and each arrow is a
bifundamental field. The difference in ranks counts the number of
fractional D5-branes.} {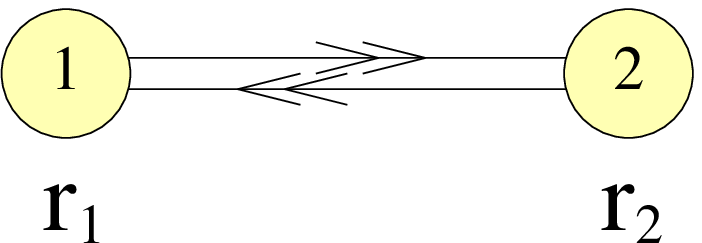}{0.7}

The theory has a superpotential of the schematic form
\eqn\wks{W =
h ~{\rm tr} \left( \epsilon_{ij} \epsilon_{kl} A^i B^k A^j B^l
\right)}
where the $A$'s and $B$'s represent the bifundamentals,
and $i,j$ ($k,l$) are global SU(2) indices for the flavor symmetry
rotating the $A$'s ($B$'s). If one chooses $r_1 = N, ~r_2=N+M$
with $N >> M$, this theory enjoys an RG cascade described in
\KlebanovHB.

A simple generalization of this singularity can be obtained by
taking a $\IZ_n$ quotient.  One class of such orbifolds \UrangaVF\
is described by the equation \eqn\orbcon{ (xy)^n = zw } in
$\IC^4$. Branes at this singularity are governed by the quiver
field theory with $2n$ nodes and bifundamentals of both
chiralities going between adjacent nodes, as in \tfig\figtwo:

\ifigure\figtwo{The quiver diagram of the orbifolded model for
$n=3$.}{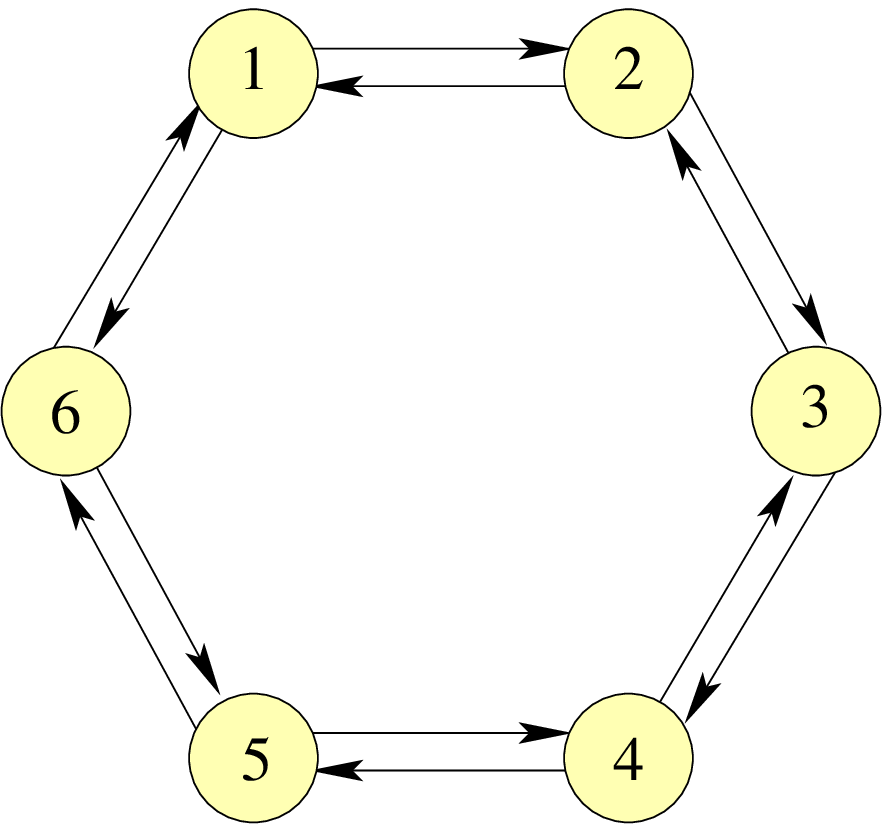}{1.5} \

\noindent This theory enjoys a quartic superpotential of the form
\eqn\wis{ W = \sum_i (-1)^i h Q^{(i)} Q^{(i+1)} \tilde Q^{(i+1)}
\tilde Q^{(i)} } where $Q^{(i)}$ corresponds to the arrow pointing
between node $i$ and node $i+1$ and so forth, and the contraction
on gauge indices is the obvious one. If one is interested in a
cascading solution with all ranks deviating from some large $N$ by
finite amounts in the UV, one can think of the matter fields $Q$
as having dimension $3/4$, and the parameter $h$ as being (to
leading order in $1/N$) dimensionless. In the IR, where the
solution departs significantly from its approximation by the CFT
which exists when all ranks are equal, it is more appropriate to
think of $h$ as being the inverse of some mass scale which is
larger than any of the gauge theory dynamical scales
$\Lambda^{(i)}$.

The class of quivers depicted in \figtwo\ was studied in
\refs{\ArgurioNY, \ArgurioQK}, as a simple home for metastable
supersymmetry breaking vacua in string theory.  The subquiver
which was relevant there, and which should arise in the IR limit
of any proposed UV completion, is shown in \tfig\figthree. For
appropriate choices of the dynamical scales $\Lambda^{(i)}$
associated with nodes $3$, $4$ and $5$, this theory was argued to
give rise (at low-energy) to a SUSY QCD theory with $N_f = N_c +
1$ slightly massive flavors, and hence to admit metastable vacua
analogous to those described in \IntriligatorDD.  (A similar
construction was described in \KitanoXG).

\ifigure\figthree{The gauge theory of interest, which can be
engineered in any of the geometries above with $n \geq
3$.}{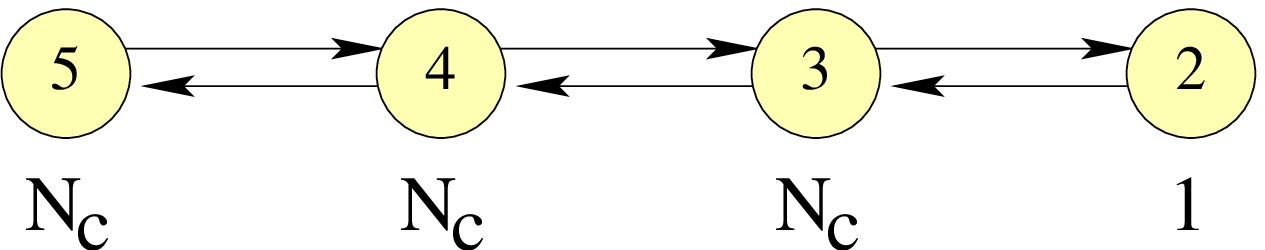}{0.7}

The only relevant point for us is that one of these non-vanishing
masses arises from a stringy instanton, wrapping the (unoccupied)
node to the right of the $U(1)$ factor in the full quiver diagram.
This instanton was argued, in suitable circumstances, to give rise
to a mass term \eqn\winst{W = \cdots + m Q^{(2)} \tilde Q^{(2)} }
for the $N_c + 1$'st quark flavor of the gauge group at node $3$.

\newsec{The stringy instanton}

First, we should specify the circumstances in which the instanton
is believed to contribute.  In the absence of the space-filling
branes, the instanton would break half of the supersymmetry in an
${\cal N}=2$ supersymmetric Calabi-Yau compactification.  Hence,
one could obtain four fermion zero modes by acting on the
instanton solution with the broken supercharges. These arise in
the sector of open strings stretching from the instanton to
itself.

However, to generate a contribution to the space-time
superpotential, there should only be two fermion zero modes (that
aren't soaked up in the integral over instanton collective
coordinates).  Although the other branes present in the quiver
gauge theory do break the space-time supersymmetry to ${\cal
N}=1$, they do not apparently lift the extra two fermion zero
modes. Therefore, to obtain a contribution to the superpotential,
one should either:

\noindent a) Consider a slightly modified configuration, where the
instanton wraps a node that also intersects an orientifold plane
\refs{\BianchiFX,\ArgurioVQ,\BianchiWY,\IbanezRS}.  The
orientifold projection can eliminate precisely half of the fermion
zero modes in the relevant open string sector, leaving the needed
two.

\noindent b) Consider a full compactification with background
fluxes (and perhaps other ingredients) in the vicinity of the
instanton. On general grounds, one would expect that in such
cases, the brane could locally detect that the background
preserves only ${\cal N}=1$ supersymmetry, and would have only two
zero modes generated by acting with broken supercharges.  It is
important to study the precise circumstances in which this
happens, of course.  Results in this direction, for instantons
which do not intersect space-filling D-branes, are implicit in
\WittenBN\ (where the geometry of F-theory reduces the number of
zero modes on certain D3-instantons to the required two), and are
generalized to models with flux in \refs{\GorlichQM
\TripathyHV\SaulinaVE\KalloshGS\BergshoeffYP-\LustCU}.

We shall proceed with option a).  In fact, a class of orientifold
models which leave the gauge theory on the space-filling branes
unmodified, while allowing the instanton to intersect an O-plane,
was already described in \refs{\ArgurioQK,\FrancoII}.

It is easier to describe the relevant geometries in the T-dual
type IIA picture.  Recall that e.g. the quiver in \figthree\ can
be T-dualized to a type IIA brane configuration.  The occupation
numbers of the nodes map to numbers of D4 branes stretching on a
circle in the $x_6$ direction between NS 5 branes stretched in the
012345 directions and NS 5' branes stretched in the 012389
directions -- the NS and NS' branes alternate as one goes around
the circle. Our configuration is shown in \tfig\figfour.

\ifigure\figfour{Type IIA T-dual description of the type IIB
quiver in \figthree.} {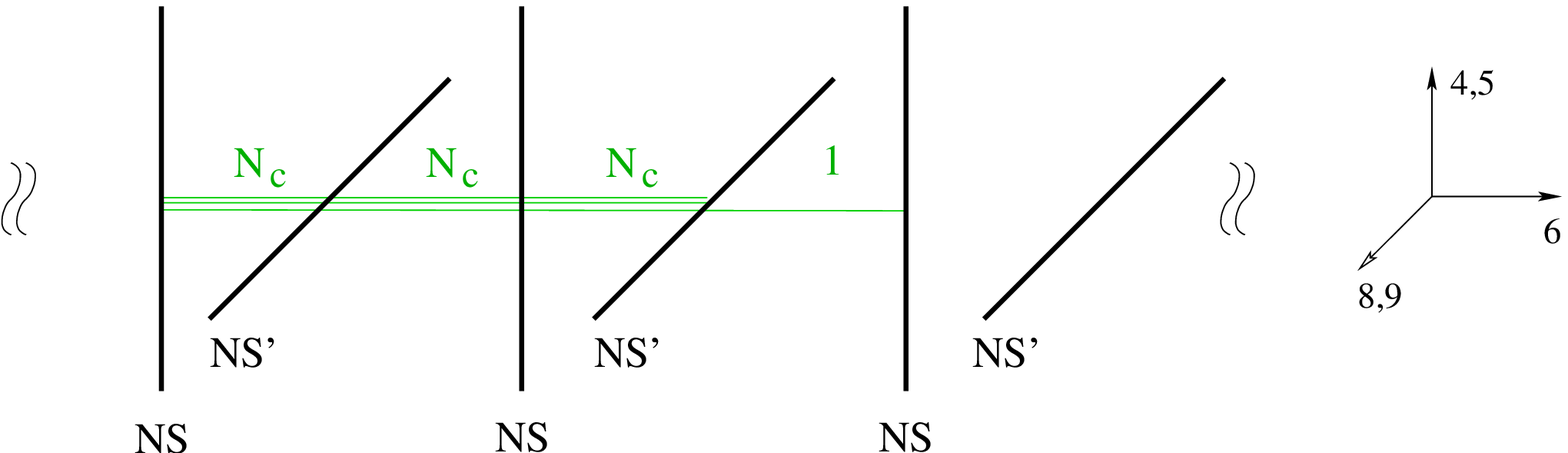}{1.7}

To make a quiver gauge theory which keeps the physics at our
occupied nodes unchanged, while allowing the D-brane instanton at
node $1$ to acquire the required zero mode content, we introduce
orientifolds as in \tfig\figfive. This configuration can be
obtained by orientifolding the $n=5$ model in an obvious way.  The
O6$^-$-planes extend along the 01237 directions, and lie at a 45
degree angle with respect to the 45 and 89 planes.

\ifigure\figfive{An orientifold which preserves our gauge theory
and contains the required instantons.}{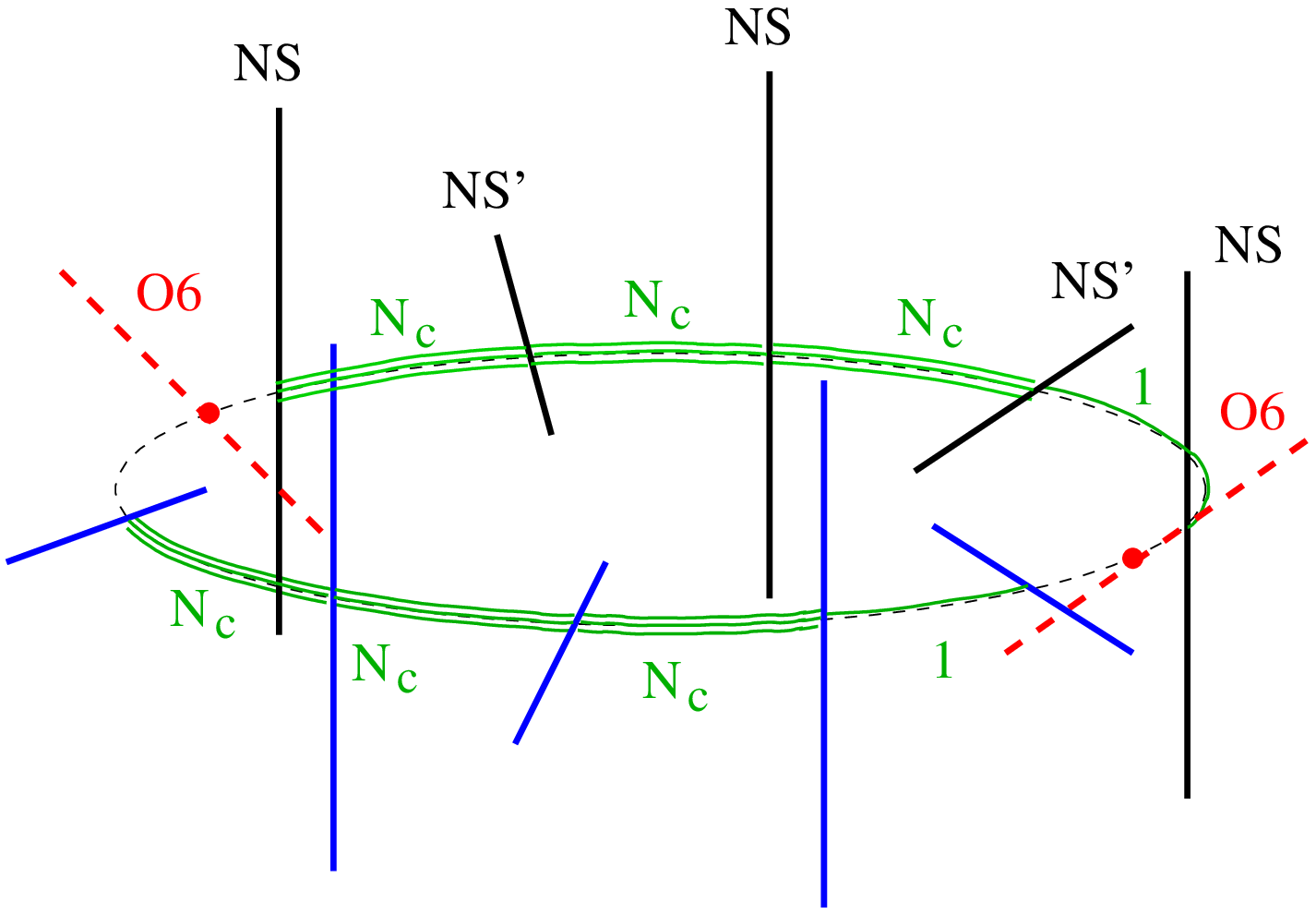}{2.0}

In this geometry (or the generalization including any number of
additional $SU(N_i)$ nodes between the O-planes), the D-instanton
wrapping the node to the right of the $SU(1)$ node, i.e. wrapping
node $1$, has $SO(1)$ worldvolume gauge group (while space-filling
branes occupying the same node would have symplectic gauge
groups). This $SO$ projection lifts the extra two fermion zero
modes, so this D-instanton can potentially contribute to the
space-time superpotential. It has in addition collective
coordinates (``Ganor strings'' \GanorPE) which stretch to node
$2$, as in the extended quiver diagram shown in \tfig\figsix.

\ifigure\figsix{The extended quiver diagram, including a node for
the Euclidean D-brane.}{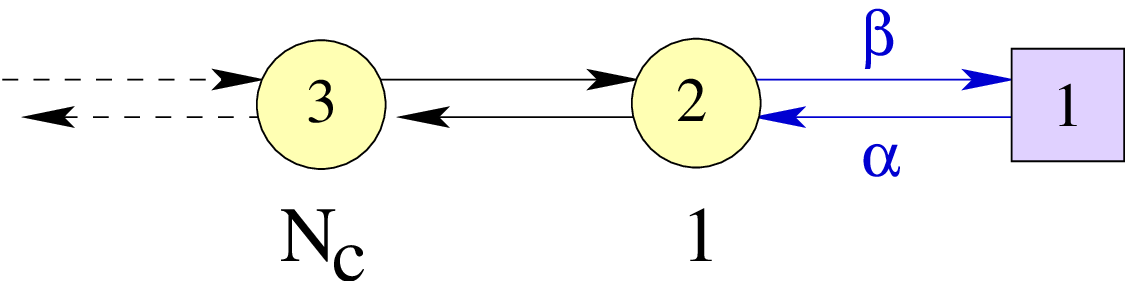}{0.9}

The instanton action is expected to contain a quartic coupling
\eqn\instag{S = ...+ \alpha Q^{(2)} \tilde Q^{(2)} \beta ~.}
After integrating out the $\alpha, \beta$ collective coordinates,
one will be left in the space-time theory with a mass term \winst,
with $m$ suppressed by the exponential of the area $A$ of node $1$.
Note that this effect only depends on nodes $1$, $2$ and $3$, and
should arise in any theory containing these nodes (as depicted in
\figsix).

In the next sections we will derive the same superpotential \winst\
in an alternative way, by embedding the gauge theory in a
renormalization group cascade. In this picture, the effect of the
stringy instanton is reproduced by a gauge theory computation, and
the exponential suppression of $m$ due to the instanton action
becomes exponential suppression by the dynamical scales of
asymptotically-free gauge group factors (which confine and disappear
at low energies).

\lref\SeibergPQ{
  N.~Seiberg,
  ``Electric - magnetic duality in supersymmetric non-Abelian gauge theories,''
  Nucl.\ Phys.\  B {\bf 435}, 129 (1995)
  [arXiv:hep-th/9411149].
}
\lref\StrasslerQS{
  M.~J.~Strassler,
  ``The duality cascade,''
  arXiv:hep-th/0505153.
}
\lref\IntriligatorNE{
  K.~A.~Intriligator and P.~Pouliot,
  ``Exact superpotentials, quantum vacua and duality in supersymmetric
  $Sp(N_c)$
  gauge theories,''
  Phys.\ Lett.\  B {\bf 353}, 471 (1995)
  [arXiv:hep-th/9505006].
}
\lref\SeibergBZ{
  N.~Seiberg,
  ``Exact Results On The Space Of Vacua Of Four-Dimensional Susy Gauge
  Theories,''
  Phys.\ Rev.\  D {\bf 49}, 6857 (1994)
  [arXiv:hep-th/9402044].
}

\def\bar{\overline}
\def\hat{\widehat}
\def\tilde{\widetilde}
\def\frac#1#2{{#1\over#2}}

\def\inbar{\,\vrule height1.5ex width.4pt depth0pt}
\def\IC{\relax\hbox{$\inbar\kern-.3em{\rm C}$}}
\def\IR{\relax{\rm I\kern-.18em R}}
\def\IP{\relax{\rm I\kern-.18em P}}

%
%

%
\catcode`\@=11
\def\slash#1{\mathord{\mathpalette\c@ncel{#1}}}
\overfullrule=0pt

\def\II{{\cal I}}

\def\underrel#1\over#2{\mathrel{\mathop{\kern\z@#1}\limits_{#2}}}

\catcode`\@=12

%

\def\det{{\rm det}}
\def\tr{{\rm tr}}

\def\det{{\rm det}}


\def\tQ{{\tilde Q}}
\def\tq{{\tilde q}}
\def\tM{{\tilde M}}
\def\hN{{\hat N}}
\def\hM{{\hat M}}
\def\tB{{\tilde B}}
\newsec{The cascade and its orientifold}

\subsec{The $U(N_i)$ cascade}

In this subsection we briefly review the cascade in the quiver
theories of \S2. At some energy scale, these theories can be
described as a $\prod_{i=1}^{2n} SU(N_i)$ gauge theory, with
chiral multiplets $Q^{(i)}$ and $\tQ^{(i)}$ ($i=1,\cdots,2n$) in
the ${\bf (N_i, \bar{N_{i+1})}}$ and ${\bf (\bar{N_i}, N_{i+1})}$
representations, respectively ($i+1$ is defined modulo $2n$). The
quiver diagram for this theory is depicted in Figure 2; it is a
generalization of the original cascade of Klebanov and Strassler
\KlebanovHB\ which arises for $n=1$. The theory has an effective
superpotential (in an arbitrary normalization of the fields) of
the form
\eqn\unsup{W = \sum_{i=1}^{2n} (-1)^i Q^{(i)} Q^{(i+1)} \tQ^{(i+1)}
\tQ^{(i)},}
with the obvious contractions of indices. Throughout this section we
will ignore numerical factors which will not be important for our
considerations.

In this subsection we will analyze a generic high-energy step of
the cascade, assuming that all the $N_i$'s are large and
comparable. Each gauge group in this theory has some strong
coupling scale $\Lambda^{(i)}$. The general analysis of this
theory is very complicated, but the analysis becomes simple when
there are large ratios between all the scales $\Lambda^{(i)}$;
then we can analyze the dynamics of each gauge theory separately
as we go down in energy, ignoring the dynamics of the other gauge
groups. A particularly simple ordering is
\eqn\niceorder{\Lambda^{(1)} \gg \Lambda^{(3)} \gg \cdots \gg
\Lambda^{(2n-1)} \gg \Lambda^{(2)} \gg \cdots \gg \Lambda^{(2n)}.}
In the order \niceorder\ the $SU(N_1)$ theory, which has $N_f =
N_2 + N_{2n}$ becomes strongly coupled first. Ignoring all other
interactions (including the superpotential interactions which are
assumed to be small at the scale $\Lambda_1$), the infrared
dynamics of this theory is the same as that of its Seiberg dual
\SeibergPQ, so we can use the variables of the dual theory instead
of our original variables (a detailed description of the
justification for this may be found in \StrasslerQS). The dual is
an $SU(N_f-N_1)=SU(N_2+N_{2n}-N_1) = SU({\hat N}_1)$ gauge theory,
whose degrees of freedom are quarks $q^{(1)}$, $\tq^{(1)}$,
$q^{(2n)}$ and $\tq^{(2n)}$ in the ${\bf (\hat N_1, N_2),
(\bar{\hat N_1}, \bar{N_2}), (\bar{N_{2n}}, \bar{\hat N_1}),
(N_{2n}, \hat N_1)}$ representations, respectively, and mesons
$M^{(2n,2n)}$ (in the adjoint+singlet representation of
$SU(N_{2n})$), $M^{(2,2)}$ (in the adjoint+singlet representation
of $SU(N_2)$), $M^{(2n,2)}$ (in the $\bf (N_{2n}, \bar{N_2})$
representation) and $M^{(2,2n)}$ (in the $\bf (\bar{N_{2n}}, N_2)$
representation). The superpotential of this theory, including the
relevant terms from \unsup\ (translated into the new variables),
takes the form (with obvious contractions)
\eqn\unfirstsup{\eqalign{ W & = M^{(2,2)} q^{(1)} \tq^{(1)} +
M^{(2n,2n)} q^{(2n)} \tq^{(2n)} + M^{(2,2n)} \tq^{(2n)} \tq^{(1)}
+ M^{(2n,2)} q^{(1)} q^{(2n)}\cr &+ M^{(2,2n)} M^{(2n,2)} -
M^{(2,2)} Q^{(2)} \tQ^{(2)} - M^{(2n,2n)} Q^{(2n-1)}
\tQ^{(2n-1)}.}}
The mesons $M^{(2n,2)}$ and $M^{(2,2n)}$ are massive so they can be
integrated out as we go to lower scales; this leaves a
superpotential of the form
\eqn\unfirstsupn{\eqalign{ W & = M^{(2,2)} q^{(1)} \tq^{(1)} +
M^{(2n,2n)} q^{(2n)} \tq^{(2n)} - q^{(2n)} q^{(1)} \tq^{(1)}
\tq^{(2n)}\cr &- M^{(2,2)} Q^{(2)} \tQ^{(2)} - M^{(2n,2n)}
Q^{(2n-1)} \tQ^{(2n-1)}.}}
The theory at lower scales has been modified in three important ways
(in additional to small changes in the charge assignments); instead
of an $SU(N_1)$ gauge theory we have an $SU(N_2+N_{2n}-N_1)$ theory,
we have two additional sets of adjoint+singlet fields for the
$SU(N_2)$ and $SU(N_{2n})$ nodes, and these fields have trilinear
superpotential couplings to all the quarks surrounding them, which
replace the two quartic couplings centered on these two nodes.

We can now perform a similar analysis for the $SU(N_3)$ node,
which will become dualized to an $SU(N_2+N_4-N_3)$ theory. Most of
the analysis is the same, except that one of the trilinear
couplings in \unfirstsupn\ now becomes a mass term for $M^{(2,2)}$
and for the field with the same quantum numbers arising as a meson
in the $SU(N_3)$ theory. Thus, these fields can also be integrated
out, and after this step we are left with no adjoint for the
$SU(N_2)$ theory, and its quartic coupling is reinstated (by
integrating out the massive mesons), with an opposite sign
compared to the original quartic coupling in \unsup. Thus, after
this step we get a theory in which the $SU(N_{2n})$ and $SU(N_4)$
nodes are modified (by adding adjoints and their associated
couplings) and the other nodes are not. If we now continue to
perform a similar analysis for the $SU(N_5)$, $SU(N_7)$, $\cdots$,
$SU(N_{2n-1})$ theories, we eventually end up with a theory of the
same form as we started with; in the last step we have two sets of
adjoint fields becoming massive. The only differences (except for
the overall sign of the superpotential) are that all fundamentals
of the odd nodes have become anti-fundamentals (and vice versa),
and the ranks of the odd nodes have been modified to $SU(N_i) \to
SU(N_{i-1}+N_{i+1}-N_i)$. The total rank of all groups together
has been reduced by
\eqn\ncascade{N_{cascade} = 2 \sum_{i=1}^n (N_{2i-1}-N_{2i});}
this must be positive in order to have a cascade and to be
consistent with \niceorder.

Obviously, we can now perform a similar analysis for all the even
nodes. After the additional $n$ steps we again go back to a theory
of the same form, with the total rank of all groups again reduced
by the same amount $N_{cascade}$. The field representations and
the overall sign of the superpotential are now the same as in the
original theory we started from. We can continue cascading down in
energy, going back to the same theory (with reduced ranks) after
every $2n$ steps. Eventually, the ranks become small enough so
that the group that becomes strongly coupled goes outside the
range $N_c+2\leq N_f < 3N_c$ where we can use Seiberg duality as
above. One then has to check in detail what happens next; in some
cases one obtains confinement in the IR theory.

Finally, let us discuss the global symmetries. There are $(2n)$
obvious vector-like $U(1)$ symmetries, acting on each pair
$Q^{(i)}$, $\tQ^{(i)}$.
In addition, we have in the classical theory a $U(1)_R$ symmetry
(with $(Q^{(i)},\tilde{Q}^{(i)})$ both having charge $1/2$) and a
single axial $U(1)_A$ symmetry which is preserved by the
superpotential \unsup, under which $(Q^{(i)}, \tilde{Q^{(i)}})$
both have charge $(-1)^i$. For generic values of the $N_i$ which
lead to a cascade, both of these symmetries are anomalous. Note
that these two $U(1)$ symmetries do not remain fixed along the
cascade; if we make some charge assignment at some energy scale,
we get a different charge assignment after going through a cascade
step (with explicit factors of $\Lambda$'s in the superpotential
to fix its charge). However, we can always rescale the
bi-fundamentals at each step of the cascade by some factors of
$\Lambda$'s to go back to the simpler charge assignments described
above.

\subsec{The orientifolded cascade}

We will now discuss a particular orientifold of this cascade,
which was described in the appendix of \ArgurioQK\ and reviewed in
\S3. The orientifold acts as a reflection on the quiver diagram
generalizing \figtwo\ to arbitrary $n$. The groups $SU(N_1)$ and
$SU(N_{n+1})$ are identified with themselves with a symplectic
projection (this is only consistent when $N_1$ and $N_{n+1}$ are
both even), while the group $SU(N_2)$ is identified (up to an
outer automorphism exchanging fundamentals and anti-fundamentals,
due to the orientation reversal) with $SU(N_{2n})$, $SU(N_3)$ with
$SU(N_{2n-1})$, and so on (up to $SU(N_{n})$ which is identified
with $SU(N_{n+2})$). All in all, we end up with a gauge group
\eqn\origroup{USp(N_1)\times SU(N_2)\times SU(N_3)\times
\cdots\times SU(N_n)\times USp(N_{n+1}).}
The matter content still includes bi-fundamentals $Q$ and
anti-bi-fundamentals $\tQ$ between all adjacent group factors in
\origroup. The superpotential is similar to \unsup, except that on
the two edges we have additional quartic terms coming from the
projection of \unsup\ using the identification described above :
\eqn\orisup{W = Q^{(1)} Q^{(1)} \tQ^{(1)} \tQ^{(1)} +
\sum_{i=1}^{n} (-1)^i Q^{(i)} Q^{(i+1)} \tQ^{(i+1)} \tQ^{(i)} +
(-1)^{n+1} Q^{(n+1)} Q^{(n+1)} \tQ^{(n+1)} \tQ^{(n+1)},}
where in the first and last terms the two $Q$'s (and the two
$\tQ$'s) are contracted in the $USp$ group, so that they give an
anti-symmetric tensor of the adjacent $SU$ group.

The cascade in this theory is very similar to the previous one,
except for the steps involving the ``edge nodes''. Again, let us
assume for simplicity that
\eqn\orinice{\Lambda^{(1)} \gg \Lambda^{(3)} \gg \cdots \gg
\Lambda^{(2)} \gg \Lambda^{(4)} \gg \cdots.}
In the first step of the cascade we then need to perform a Seiberg
duality on the $USp(N_1)$ group which has $N_f=2N_2$ flavors. This
proceeds as described in \IntriligatorNE, and turns the gauge
group into $USp(2N_2-N_1-4)$. The field content involves new
bi-fundamental and anti-bi-fundamental quarks $q^{(1)}$ and
$\tq^{(1)}$, and mesons $M^{(2,2)}$ in the (adjoint+singlet)
representation of $SU(N_2)$, $M^{A}$ in the anti-symmetric tensor
representation of $SU(N_2)$ and $M^{\bar A}$ in the conjugate
anti-symmetric representation. The superpotential, including the
terms coming from the duality transformation as well as the
relevant terms of \orisup, is
\eqn\newsup{W = M^{(2,2)} q^{(1)} \tq^{(1)} + M^{A} \tq^{(1)}
\tq^{(1)} + M^{\bar A} q^{(1)} q^{(1)} + M^A M^{\bar A} -
M^{(2,2)} Q^{(2)} \tQ^{(2)}.}
The fields $M^A$ and $M^{\bar A}$ are massive and can be
integrated out, leading to a quartic term $-q^{(1)} q^{(1)}
\tq^{(1)} \tq^{(1)}$. The quartic terms involving the second node
have been replaced by having an adjoint+singlet field $M^{(2,2)}$
with trilinear couplings, just as in the discussion of the
previous subsection.

If we now dualize the group $SU(N_3)$, then again this will give a
mass to $M^{(2,2)}$ (and regenerate the quartic couplings
involving this mode), and generate a new adjoint field $M^{(4,4)}$
with trilinear couplings. Continuing along the cascade, if $n$ is
even we finish the odd steps by dualizing $USp(N_{n+1})$, while if
$n$ is odd we finish by dualizing $SU(N_n)$. In both cases, after
this we return to a theory of the same form (up to some signs and
conjugations) as the original theory. The total rank is reduced by
\eqn\ncasone{N_{cascade} = N_1 + 2 - 2 N_2 + 2 N_3 - \cdots - 2
N_n + N_{n+1} + 2}
when $n$ is even, and by
\eqn\ncastwo{N_{cascade} = N_1 + 2 - 2 N_2 + 2 N_3 - \cdots + 2
N_n - N_{n+1}}
when $n$ is odd.

As before, we can now perform a similar cascade involving the even
nodes. The duality of $SU(N_2)$ does not give an adjoint for the
adjacent $USp$ group (since the superpotential makes this
massive), but it does give an adjoint for the adjacent $SU$ group.
As we go down the cascade, eventually all the adjoints become
massive, and we go back to the a theory with the same form as the
original theory, just as in the previous case. We can then
continue cascading until the ranks become too small to perform
further Seiberg dualities, and a different analysis is required in
the IR.

Finally, let us describe the global symmetries in this case. We
now have $(n+1)$ vector-like $U(1)$ global symmetries. There is
still a classical $U(1)_R$ symmetry (with all bi-fundamentals
having charge $1/2$), but there is no longer any axial $U(1)$
consistent with \orisup. For generic values of the $N_i$ (for
which $N_{cascade} \neq 0$, as required in order to have a
cascade) the $U(1)_R$ symmetry is anomalous. As before, this
$U(1)_R$ is not invariant under the cascade, unless we rescale the
bi-fundamentals by appropriate powers of $\Lambda$'s.

\newsec{The IR physics of the cascade: non-perturbative mass generation}

We will now discuss the IR physics of the cascade in two special
cases. We assume that the cascade proceeds as in the previous
section, until we need to perform a duality on the $USp(N_1)$ node
but its rank is not large enough.

The first case we discuss is the case discussed in section 3, in
which we want to end up with $USp(0)\times SU(1)\times \cdots$.
For this we need to start higher up in the cascade with $N_1 =
2N_2 - 4$, and with $N_4 = N_3 + 1$. The $USp(N_1)$ theory now has
$2N_2 = N_1 + 4$ flavors. In this case the low-energy dynamics of
this theory does not involve any modification of the classical
moduli space for the mesons $M^{(2,2)}$, $M^A$ and $M^{\bar A}$.
Rather, there is an effective superpotential implementing the
classical constraints. If we denote the full anti-symmetric meson
matrix of the $USp(N_1)$ theory by ${\cal M}$ (${\cal M}_{ij} =
M^A_{ij}$, ${\cal M}_{i(j+N_2)}=M^{(2,2)}_{ij}$, ${\cal
M}_{(i+N_2)(j+N_2)} = M^{\bar A}_{ij}$ for all $i,j=1,\cdots,
N_2$), then the low-energy superpotential is \IntriligatorNE\ of
the form (we will ignore constants and powers of $\Lambda$'s in
all our expressions here, as before)
\eqn\spnoconstr{W = {\rm Pf}({\cal M})
.}
The other couplings of the mesons are the same as the last two
terms in \newsup\ above, which come from the quartic
superpotential. Since the fields $M^A$, $M^{\bar A}$ are massive
we can integrate them out, and end up with a superpotential
\eqn\spsuppot{W = \det(M^{(2,2)})
- M^{(2,2)} Q^{(2)} \tQ^{(2)}.}

As described above, the next stage in the cascade involves
dualizing the $SU(N_3)$ group to a
$SU(\hN_3)=SU(N_4+N_2-N_3)=SU(N_2+1)$ group. In this step we
replace $Q^{(2)} \tQ^{(2)}$ by a new meson field $\tM^{(2,2)}$,
and the relevant terms in the superpotential become
\eqn\spsuppotn{W = \det(M^{(2,2)})
-
M^{(2,2)} \tM^{(2,2)} + \tM^{(2,2)} q^{(2)} \tq^{(2)}.}
The fields $M^{(2,2)}$ and $\tM^{(2,2)}$ are now massive;
integrating them out means that we can replace $M^{(2,2)}$ in
\spsuppotn\ by $q^{(2)} \tq^{(2)}$, where the $SU(\hN_3)$ indices
are contracted and the $SU(N_2)$ indices are not; we will denote
this matrix by $(q\tq)^{(2,2)}$.

We can now continue down the cascade, dualizing all the odd nodes.
Next, we need to analyze the low-energy dynamics of the $SU(N_2)$
node. This node has $N_f={\hat N}_3 = N_2+1$ coming from the
$q^{(2)}$ and $\tq^{(2)}$ flavors, so its low-energy description
is in terms of mesons and baryons, with an effective
superpotential imposing the classical constraints relating the
mesons and the baryons \SeibergBZ. The mesons $\hM^{(3,3)}$ are an
adjoint+singlet of $SU(\hN_3)$, while the baryons $B^{(2)}$ and
anti-baryons ${\tilde B}^{(2)}$ are in the fundamental and
anti-fundamental representations. Like any other gauge-invariant
chiral operator, we can write $\det((q\tq)^{(2,2)})$ in terms of
these mesons and baryons. In fact, there are several ways to do
this. On one hand it is equal to the subdeterminant of the mesons,
which is a polynomial of rank $N_2$ in the traces of the (${\hat
N}_3\times {\hat N}_3$) meson matrix $\hM^{(3,3)}$, of the form
\eqn\polym{\eqalign{{\rm subdet}(M) &= N_2! \sum_{l=1}^{\infty}
\sum_{n_i=1;i=1,\cdots,l}^{N_2} {{(-1)^{l+N_2}}\over {l!}}
\delta_{\sum_i n_i, N_2} \prod_{i=1}^l {{\tr(M^{n_i})}\over n_i} \cr
&= \tr(M)^{N_2} + \cdots + (N_2-1)! (-1)^{1+N_2} \tr(M^{N_2}).}}
On the other hand, it is equal to a product of baryons $B^{(2)}
\tB^{(2)}$, with a contraction of their $SU(\hN_3)$ indices. These
two expressions are the same classically, and in this theory this
equivalence remains true also quantum-mechanically; the
superpotential which imposes that the moduli space is equal to the
classical moduli space makes these two expressions the same in the
chiral ring (namely, they are the same up to the addition of
non-chiral operators).

At low energies the $SU(N_2)$ group is gone. Making a specific
choice of writing the determinant operator $\det((q\tq)^{(2,2)})$
using baryons rather than mesons (all other choices are equivalent
in the superpotential), we obtain an effective action
\eqn\spsuppotnn{W = \hM^{(3,3)} (B^{(2)} {\tilde B}^{(2)} - q^{(3)}
\tq^{(3)}) - \det(\hM^{(3,3)}) + B^{(2)} \tB^{(2)}.}
%

Next, we want to perform a duality on the $SU(N_4)$ theory. This
turns $q^{(3)} \tq^{(3)}$ into a new meson $M^{(3,3)}$, that
couples trilinearly to new quarks $B^{(3)}$ and $\tB^{(3)}$. The
equation of motion of $M^{(3,3)}$ now relates $\hM^{(3,3)}$ to
bilinears of these quarks. Integrating out all the massive fields
reduces the superpotential to
\eqn\finalsup{W = B^{(2)} B^{(3)} \tB^{(3)} \tB^{(2)} - \det(B^{(3)}
\tB^{(3)}) + B^{(2)} \tB^{(2)},}
%
where in the second term the quarks are contracted to give an
adjoint+singlet of $SU({\hat N}_3)$.

The quiver theory that we ended up with is the same as the one
described in Figure 6 (identifying ${\hat N}_3 = N_c$). The
superpotential we find is also the same as we found there (up to
the irrelevant determinant operator in \finalsup). In the
derivation from the cascade it is not obvious that the mass term
in \finalsup\ is related to a one-instanton effect in the
``$USp(0)$'' gauge group that we get at the end. However, it is
easy to verify that such a one-instanton term would have the same
quantum numbers (=the same anomalous R-charge) as the term we
got\foot{This would not be the case for the cascade involving only
$SU$ groups that we discussed in \S4.1, consistent with the fact
that no mass is generated by the stringy instanton in that case.}.
Note that a one-instanton term in a specific gauge group in a
cascading theory does not translate into a one-instanton term
higher up in the cascade. This is evident from the computation in
this section, in which the stringy instanton effect arises not
from instantons but from strongly coupled dynamics of the gauge
groups higher up in the cascade. Even though a single Seiberg
duality transforms a one-instanton term into a one-instanton term,
this is not true of the full cascade, which involves multiple
Seiberg dualities on all the nodes (including nodes which are
``flavor'' nodes from the point of view of the instanton).

Similarly, we can analyze the case of $N_1 = 2 N_2 - 2$, which
formally ends up after the cascade step with ${\hat N}_1=-2$. In
this case, the low-energy dynamics of the $USp(N_1)$ theory leads
to a quantum modified moduli space \IntriligatorNE. So, instead of
the terms involving $\det(M^{(2,2)})$ in the superpotential, we
get constraint terms $\lambda (\det(M^{(2,2)}) - 1)$ with a
Lagrange multiplier $\lambda$. All the later dualities are not
modified, so at the end of the cascade step in this case we obtain
the same superpotential as \finalsup, but with the last term
replaced by $\lambda (B^{(2)} \tB^{(2)} - 1)$.

\medskip

\centerline{\bf{Acknowledgements}}

We would like to thank R. Argurio, M. Bertolini, B. Florea, S.
Franco, N. Seiberg, E. Silverstein, A. Tomasiello and A. Uranga
for helpful discussions. The work of OA is supported in part by
the Israel-U.S. Binational Science Foundation, by a center of
excellence supported by the Israel Science Foundation (grant
number 1468/06), by the European network HPRN-CT-2000-00122, by a
grant from the G.I.F., the German-Israeli Foundation for
Scientific Research and Development, and by a grant of DIP (H.52).
The research of SK was supported in part by NSF grant PHY-0244728,
and in part by the DOE under contract DE-AC03-76SF00515.

\listrefs

\end